\documentclass{elsart}
\usepackage{graphics}

\def\etal{{\it{}et~al.}}        % Macro to write "et al." in italics

\def\set#1{\lbrace#1\rbrace}

\def\psfigure#1#2{\resizebox{#2}{!}{\includegraphics{#1}}}
\begin{document}

\date{\today}
\journal{Proc. R. Soc. London B}

\begin{frontmatter}
\title{Effects of neutral selection on\\
the evolution of molecular species}
\author{M. E. J. Newman}
\address{Santa Fe Institute, 1399 Hyde Park Road, Santa Fe, NM 87501.
U.S.A.}
\author{Robin Engelhardt}
\address{Center for Chaos And Turbulence Studies, Dept.\ of Chemistry,\\
University of Copenhagen, Universitetsparken 5, Copenhagen \O, Denmark}
\begin{abstract}
  We introduce a new model of evolution on a fitness landscape possessing a
  tunable degree of neutrality.  The model allows us to study the general
  properties of molecular species undergoing neutral evolution.  We find
  that a number of phenomena seen in RNA sequence-structure maps are
  present also in our general model.  Examples are the occurrence of
  ``common'' structures which occupy a fraction of the genotype space which
  tends to unity as the length of the genotype increases, and the formation
  of percolating neutral networks which cover the genotype space in such a
  way that a member of such a network can be found within a small radius of
  any point in the space.  We also describe a number of new phenomena which
  appear to be general properties of neutrally evolving systems.  In
  particular, we show that the maximum fitness attained during the adaptive
  walk of a population evolving on such a fitness landscape increases with
  increasing degree of neutrality, and is directly related to the fitness
  of the most fit percolating network.
\end{abstract}
\end{frontmatter}

\section{Introduction}
\label{intro}
Biological molecules such as proteins and RNAs undergo evolution just as
organisms do, selected for their ability to perform certain functions by
the reproductive success which that ability imparts on their hosts.  It is
believed that many mutations of a molecule are evolutionarily neutral in
the sense that they do not change the fitness of the molecule to perform
the function for which it has been selected.  We have many examples of
proteins which appear to possess approximately the same conformation and to
perform the same function in different species, but which have different
sequences.  Such proteins may differ only by a single amino acid or may
have whole regions which have been substituted or inserted, or they may
even be so different as to appear completely unrelated.  A mutation is said
to be neutral if it changes a molecule into one of these functional
equivalents, leaving the viability of its host unchanged.  This idea was
first explored in detail by Kimura~\cite{Kimura55,Kimura83}.

In fact, as Ohta has pointed out~\cite{Ohta72}, it is not necessary that
the fitness of a molecule remain precisely the same under a given mutation
for that mutation to be considered neutral.  A mutation which produces a
change in fitness will cause the population sizes of the original and
mutant strains to diverge exponentially from one another over time.
However, the time constant of this exponential is inversely proportional to
the change in fitness.  Thus if the change in fitness is small, its effects
will be felt only on very long time-scales.  If these time-scales are
significantly longer than the time-scale on which mutations occur (the
inverse of the mutation rate), then the change in fitness will never be
felt.  In effect, the mutation rate places a limit on the resolution with
which selection can detect changes of fitness, so that small fitness
changes are effectively, if not precisely, neutral.

It is possible that the concept of neutral selection can also be applied to
the evolution of entire organisms.  Certainly there are changes possible in
an organism's genome which have no immediate effect on its reproductive
success, or which produce an effect sufficiently small that selection
cannot detect it on the available time-scales.  In this paper we will
primarily use the language of molecular evolution, but the reader should
bear in mind that the ideas described may have wider applicability.

Despite the long history of the idea, many aspects of neutral evolution are
still not well understood.  In particular, we have very little idea of the
general behaviours that can be expected of systems (molecules or organisms)
with a significant degree of evolutionary neutrality.  The primary reason
for this gap in our understanding is that, despite many decades of hard
work, we still have a rather poor idea of the way in which genomic
sequences map onto molecular structures and hence onto a fitness measure.
In the case of entire organisms the equivalent problem is that of
calculating the genotype-phenotype mapping, which is even less well
understood.  One simple case in which neutral evolution has been
investigated in some detail is that of RNA
structure~\cite{SFSH94,HSF96,GGSRWHSS96a,GGSRWHSS96b}, although
calculations so far are limited to secondary structures, and even these
cannot be calculated with any reliability, so that these studies should be
taken more as a qualitative guide to the behaviour of systems undergoing
neutral selection than an accurate representation of the real world.  The
trouble with this approach however is that RNAs are not a sufficiently
general model that the results gained from their study can be applied to
other systems, such as protein evolution or the evolution of whole
organisms.

At the other extreme, studies have been performed of extremely simple
mathematical models of neutral evolution in the context of genetic
algorithms~\cite{Mitchell96,PBS94}.  An example is the ``Royal Road''
genetic algorithm studied by van Nimwegen~\etal~\cite{NCM97a,NCM97b}.
These models possess highly artificial fitness functions chosen
specifically to show a high degree of neutrality whilst at the same time
being simple enough to yield to analytic methods.  Like RNA secondary
structures, these models have given us some insight into the type of
effects we may expect neutral evolution to produce, but, like RNAs, they
are not sufficiently general to be sure that these insights apply to other
systems as well.

In this paper, therefore, we propose a new mathematical model of neutral
evolution.  This model is an abstract model of a genotype to fitness map in
the spirit of the Royal Road model.  This approach allows us to sidestep
the problems of incorporating the chemistry of real molecules in our
calculations and to investigate the properties of the system more quickly
and in greater detail than is possible with, for example, RNA structure
calculations.  In addition, the model is more general than either the Royal
Road fitness function or the RNA sequence-structure maps of
Refs.~\cite{HSF96,GGSRWHSS96a}.  In fact, it possesses regimes in which it
mimics the behaviour of both of these systems, as well as protein- and
organism-like regimes.  Because the behaviours of our model cover such a
wide range of possibilities, it seems reasonable to conjecture that generic
features of the model which span all of these regimes may be common to most
systems undergoing neutral selection.  This is the power of our model, and
these general results are the results that we will concentrate on in this
paper; we believe that the generic behaviours of our model should be
visible in the evolution of real systems such as proteins which are, as
yet, beyond our abilities to study directly.

In Section~\ref{model} we introduce our landscape model of neutral
evolution.  In Section~\ref{props} we discuss its properties and compare
these with previous results for other systems undergoing neutral evolution.
In Section~\ref{discuss} we discuss the implications of our results for
evolving molecular species.  In Section~\ref{concs} we give our
conclusions.

\section{The model}
\label{model}
Selective neutrality arises as a result of the many-to-one nature of the
sequence-structure or genotype-phenotype maps found in biological systems.
Many protein sequences, for example, map onto the same tertiary structure,
and since the fitness is primarily a function of the structure, such
sequences possess (at least approximately) the same fitness.  We wish to
construct a model of this phenomenon without resorting to actual
calculations of the structure of any particular class of molecules.  Only
in this way can we hope to create a model which is general enough to
represent the behaviours of many different such classes.  Our approach is
to employ a ``fitness landscape'' model of the type first proposed by
Wright~\cite{Wright67,Wright82} which maps sequence (or genotype) directly
to fitness.  Structures (or phenotypes) appear in our model as contiguous
sets or ``neutral networks'' of sequences possessing the same fitness.

Our model is a generalization of the $NK$ model proposed by
Kauffman~\cite{KJ91,Kauffman92}, which is itself a generalization of the
spin glass models of statistical physics~\cite{FH91}.  Consider a sequence
of $N$ loci, which correspond to the nucleotides in an RNA or to amino
acids in the case of a protein.  At each locus $i$ we have a value $x_i$
drawn from an appropriate alphabet, such as $\set{\mbox{A,C,G,U}}$ for
RNAs, or the set of 20 amino acids in the case of proteins.  We denote the
size of the alphabet by $A$.  Each locus interacts with a number $K$ of
other ``neighbour'' loci, which may be chosen at random or in any other way
we wish.  (Kauffman refers to these interactions as epistatic interactions,
though this nomenclature is strictly only appropriate to the case where we
are modelling the fitness of whole organisms.)  In the case of RNAs, bases
most often interact with one other base to form either a Watson-Crick or a
G-U pair.  Some bases have both pairing and tertiary interactions.  Some,
in the single stranded regions, have very little interaction with any
others.  Thus a value of $K=1$ might be approximately correct for
RNA.\footnote{It is possible to generalize the model to allow different
  loci to interact with different numbers of neighbours, which gives a
  behaviour more representative of true RNAs and proteins.  In this paper
  however, we will confine ourselves to the case of constant $K$ for
  simplicity.  As we will see, even with this constraint the model is still
  able to duplicate the behaviours seen in real systems.} For proteins,
which have more complex types of interactions, a higher value of $K$ may be
appropriate.

Each locus $i$ makes a contribution $w_i$ to the fitness of the sequence,
whose magnitude depends on the value $x_i$ at that locus and also on the
values at each of the $K$ neighbouring loci.  There are $A^{K+1}$ possible
sets of values for the $K+1$ loci in this neighbourhood, and hence
$A^{K+1}$ possible values of $w_i$.  Following Kauffman and
Johnsen~\cite{KJ91} we choose this set of values at random.  However,
Kauffman and Johnsen chose the values to be random real numbers in the
interval $0\le w_i<1$.  We by contrast choose them to be {\em integers\/}
in the range $0\le w_i<F$.  Thus if $F=2$ for example, each contribution
$w_i$ is either zero or one.  Now we define the fitness $W$ of the entire
sequence to be proportional to the sum of the contributions at each locus:
\begin{equation}
W = {1\over N(F-1)} \sum_i w_i.
\end{equation}
The fitness of all sequences thus falls in the range from zero to one, and
there are $NF-N+1$ possible fitness values in this range.

In the limit in which $F\to\infty$ the probability that two sequences will
possess the same fitness becomes vanishingly small, and our model therefore
possesses no neutrality and is in fact exactly equivalent to the $NK$
model.  However, when $F$ is finite the probability of two sequences
possessing the same fitness is finite, so that the model possesses
neutrality to a degree which increases as $F$ decreases.  Neutrality is
greatest when $F$ takes the smallest possible value of two.  Two sequences
with the same fitness may be equivalent either to molecules which fold into
the same conformation and perform the same function, or they may be
equivalent to molecules with different conformations but approximately the
same contribution to the reproductive success of the host organism.  The
ruggedness of the landscape is controlled by the parameter $K$, and is
largest when $K$ takes the maximal value of $N-1$~\cite{KJ91,Weinberger91}.
In the next section, we investigate the properties of the landscapes
generated by our model, and show that with the right choice of parameters
they can be used to mimic real biological systems, such as RNAs.

\section{Evolution on neutral landscapes}
\label{props}
The topology of a fitness landscape depends on the types of mutation
allowed to molecules evolving on it.  In biological evolution, point
mutations---mutations of the value at a single locus---are the most common.
In this case, a neutral network is defined to be a set of sequences which
all possess the same fitness and which are connected together via such
point mutations.  In the molecular case, we assume that closely similar
sequences have the same fitness because they fold into the same
conformation, so that these neutral networks correspond to (tertiary)
structures.  In the organismal case they correspond to
phenotypes.\footnote{One might argue that the individual fitness levels
  should correspond to structures, not the neutral networks.  However,
  there are presumably many structures which are not similar enough to be
  easily accessible from one another by point mutations, and yet which
  possess similar fitnesses, at least to the mutation-limited accuracy with
  which selection can distinguish.  Thus it seems more appropriate to draw
  a correspondence between networks and structures in the case of this
  model.}

\begin{figure}
\begin{center}
\psfigure{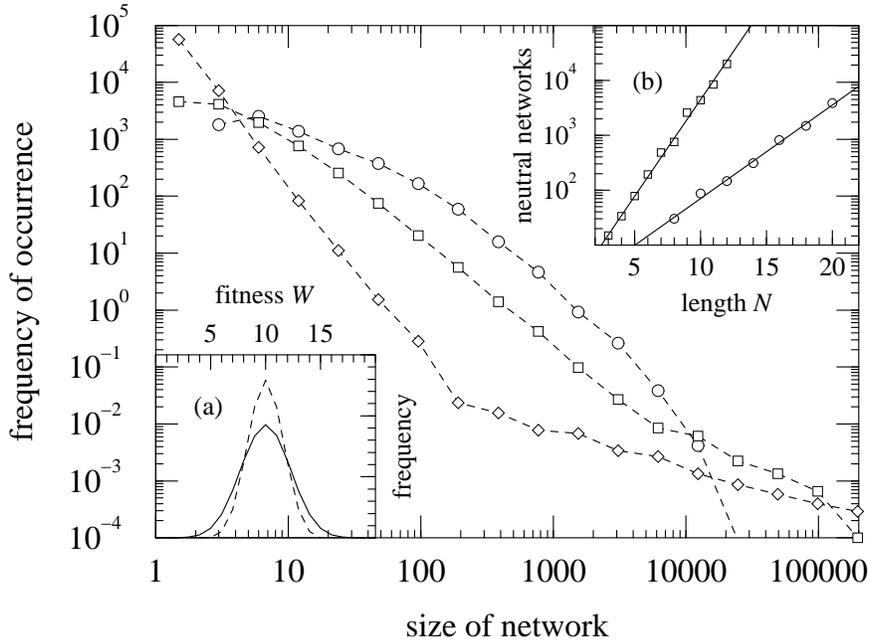}{11.5cm}
\end{center}
\caption{Histogram of the frequency of occurrence of neutral networks as a
  function of size for a particular realization of a landscape with $N=20$,
  $A=2$, $F=2$, and (circles) $K=1$, (squares) $K=2$, (diamonds) $K=5$.
  Inset (a): the frequency of occurrence of sequences as a function of
  fitness for $N=20$, $A=2$, $F=2$, and (solid line) $K=19$, (dotted line)
  $K=0$.  Inset (b): the number of neutral networks as a function of $N$
  for $K=1$, $F=2$, and (circles) $A=2$, (squares) $A=4$.
\label{fig1}}
\end{figure}

The model described in the last section possesses neutral networks of
exactly this type.  The total fitness $W$ in the model ranges from zero to
one, but the greatest number of sequences have fitness close to $W=0.5$.
(In the extreme case where $K=N-1$ the distribution of $W$ is binomial.
When $K<N-1$ it is approximately but not exactly so.  Examples of these
distributions are shown in Figure~\ref{fig1}~(a).)  We would therefore
expect the largest neutral networks to be those with fitness close to
$W=0.5$ and this is indeed what we find in practice.

Typically there are a large number of small neutral networks and a small
number of large ones.  In Figure~\ref{fig1} we show histograms of the sizes
of the neutral networks for $N=20$ and various values of $K$.  For the
RNA-like case $K=1$, the histogram appears to be convex, indicating a
distribution which falls off faster than a power law.  The same behaviour
has been in seen in RNA studies by Gr\"uner~\etal~\cite{GGSRWHSS96a}.  As
$K$ increases the distribution flattens, and by the time we reach $K=5$ it
is markedly concave.  Thus the behaviour seen in RNAs is not in this case
generic.  For some intermediate value of $K$ close to $K=2$, the
distribution appears to be power-law in form, perhaps indicating the
divergence of some scale parameter governing the distribution, in a manner
familiar from the study of critical phenomena~\cite{BDFN92}.

We find that the total number of neutral networks $S_N$ grows exponentially
as $a^N$ with sequence length.  In Figure~\ref{fig1}~(b) we show the number
of networks in our model for $K=1$, both for two-letter $\set{\mbox{G,C}}$
alphabets, and for a four-letter $\set{\mbox{A,C,G,U}}$ alphabet.  We find
that $a\approx1.5$ for the $A=2$ case and $a\approx2.3$ for the $A=4$ case.
Interestingly, Stadler and co-workers~\cite{GGSRWHSS96a,GGSRWHSS96b,SH97}
have performed the same calculations for RNA sequences using the full
secondary-structure calculation, and they also find an exponential increase
in the number of structures with sequence length with values of $a=1.6$ and
$a=2.35$ for the two- and four-letter cases respectively.  This suggests to
us that this behaviour is more general than the specific
secondary-structure map employed in the Stadler calculations.

\begin{figure}
\begin{center}
\psfigure{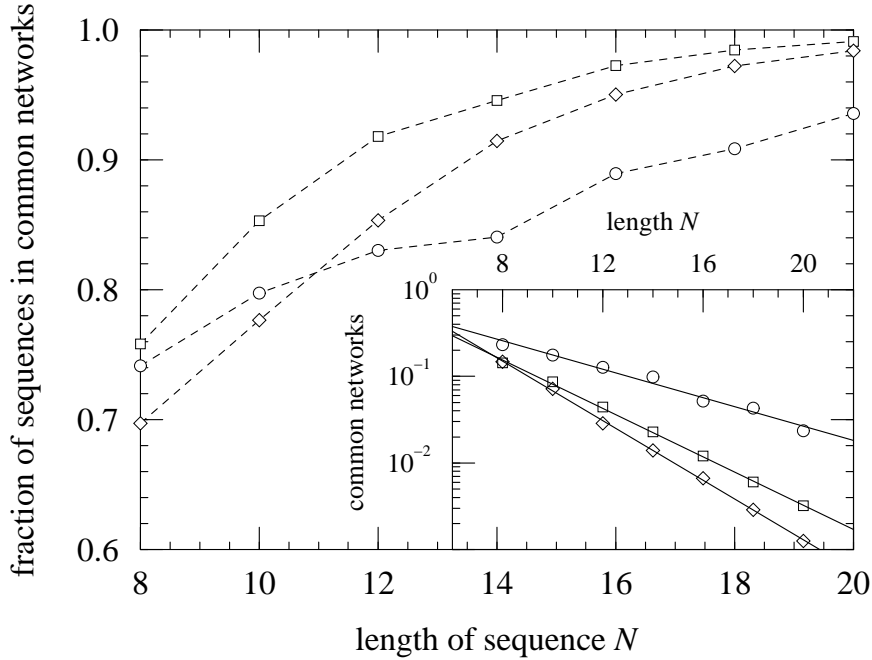}{11.5cm}
\end{center}
\caption{The fraction of all sequences which fall in common networks as a
  function of $N$ for a particular realization of the model with $A=2$,
  $F=2$, and (circles) $K=2$, (squares) $K=4$, (diamonds) $K=6$.  Inset:
  the number of common networks as a fraction of the total number of
  networks for the same landscapes.
\label{fig2}}
\end{figure}

The largest neutral networks on our landscapes {\em percolate,} which is to
say, they fill the sequence space roughly uniformly, in such a way that no
sequence is more than a certain distance away from a member of the
percolating network.  Determining which networks are percolating is not an
easy task.  We have developed two different measures to help identify
percolating networks, and we describe them further in Ref.~\cite{NE98}.
Gr\"uner~\etal~\cite{GGSRWHSS96a} introduced instead the idea of a
``common'' network, which is one which contains greater than the average
number of sequences.  We can employ this definition with our model too.  We
find that the common networks in the model form a small fraction of the
total number of networks, that fraction decreasing exponentially as $N$
increases, as shown in the inset to Figure~\ref{fig2}.  The same result is
found in RNAs~\cite{GGSRWHSS96a}.

In the main frame of Figure~\ref{fig2} we show the fraction of sequences
which fall in the common networks as a function of $N$.  As the figure
shows, our numerical studies indicate that this fraction {\em increases\/}
with sequence length, tending to one in the limit of large $N$.  Even
though the common networks form a smaller and smaller fraction of all
networks as $N$ becomes large, they nonetheless cover more and more of the
sequence space.  These results have interesting evolutionary implications:
they imply that as sequences become longer, a larger and larger majority of
structures (the small networks) are vanishingly unlikely to occur through
natural selection.  Evolution can only find the smaller and smaller
fraction of ``common'' structures.  The same conclusions have been reached
in the case of RNAs.  However, the results presented here indicate that
these conclusions are not specific to RNAs, and probably apply to most
systems undergoing neutral evolution.

\begin{figure}
\begin{center}
\psfigure{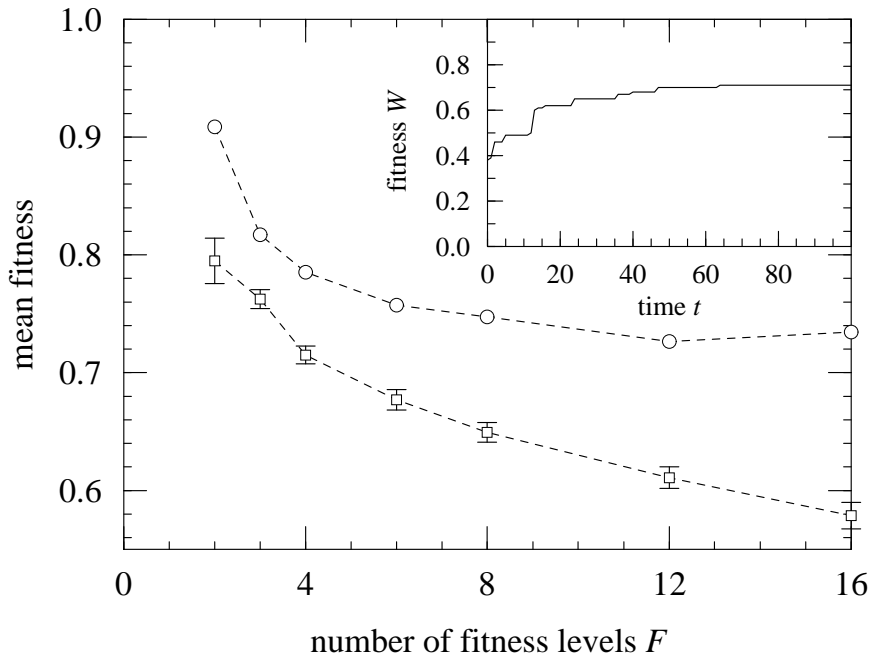}{11.5cm}
\end{center}
\caption{The maximum fitness attained by a random hill-climber averaged
  over ten simulations with $N=20$, $K=4$ and $A=2$, as a function of the
  neutrality parameter $F$ (circles).  The lower curve (squares) is the
  fitness of the most fit percolating neutral network averaged over the
  same ten runs.  Inset: the fitness of one of the hill-climbers in the
  simulation as a function of time.
\label{fig3}}
\end{figure}

Next we have examined the dynamics of populations evolving on our
landscapes.  These studies have yielded some of the most interesting
results of this work.  In their studies with $NK$ landscapes, Kauffman and
Johnsen~\cite{KJ91} made a useful approximation in representing evolving
populations by their single dominant sequence.  This approximation is only
valid in the case in which the time-scale for mutation is much longer than
the time-scale on which selection acts.  For the moment we will assume this
to be the case.  A ``random hill-climber'' is a population of this type,
represented by a single dominant strain, which tries mutations---point
mutations in the present case---until it finds one with higher fitness than
the current strain.  In this way the hill-climber performs an adaptive walk
through sequences of ever-increasing fitness until it reaches a local
fitness optimum.  To study neutral landscapes we modify this strategy so
that the hill-climber samples adjacent sequences at random until it finds
one of fitness greater than {\em or the same as\/} itself.  Such a climber
will move at random on a neutral network until it finds a mutation which
takes it to a network of higher fitness.  In the upper curve of
Figure~\ref{fig3} we show the average fitness attained by such a walker
over ten simulations on our landscapes as a function of the neutrality
parameter $F$.  Recall that neutrality increases with decreasing $F$.  As
the figure shows, the hill-climber on average finds higher fitness maxima
for higher degrees of neutrality.  In other words, neutrality helps the
population to attain a greater fitness.  This is certainly an idea which
has been entertained before in the literature, but it is lent a new
conviction when we see it emerge in the behaviour of a general model such
as this.

The lower curve on Figure~\ref{fig3} shows the fitness of the most fit
percolating network averaged over the same ten landscapes.  The curve
follows quite closely the form of the curve for the fitness of the local
maxima found by the hill-climber.  Our explanation of this result is as
follows.  The climber moves diffusively on a neutral network until it finds
a one-mutant neighbour which belongs to a network of greater fitness, at
which point it shifts to that network.  This process continues until it
reaches a non-percolating network, at which point it is confined to the
region occupied by the network and can only get as high as the local
maximum within that region.  Thus the highest fitness attainable on a
landscape with neutrality depends directly on the highest fitness at which
there are percolating networks.  Since the landscapes with the greatest
degree of neutrality also have more and fitter percolating networks, this
explains why higher fitnesses are attained on landscapes with lower values
of $F$.

The inset to Figure~\ref{fig3} shows the fitness of one of our
hill-climbers as a function of time, and we can clearly see the steps in
this function where the climber finds its way onto a neutral network of
higher fitness.  Similar steps have been seen, for example, in laboratory
experiments on the evolution of bacteria~\cite{LT94,SGL97}.  Although it
appears in the figure that nothing happens in the periods between these
jumps, it is at these times that the climber diffuses around its network,
testing new mutations to find one of higher fitness.  It is this diffusive
motion which allows us to find higher fitness sequences on landscapes with
higher degrees of neutrality.  Van Nimwegen~\etal~\cite{NCM97a} have dubbed
these periods of apparent stasis ``epochs''.  They also bear some
similarity to the palaeontological ``punctuated equilibria'' described by
Eldredge and Gould~\cite{EG72,GE93}, although there are many other possible
explanations for the periods of stasis seen in fossil evolution.

To investigate the epochs in more detail, we have performed simulations of
true populations evolving on our landscape.  In these simulations we take a
population of $M$ sequences which at each generation reproduce with
probability proportional to their fitness, in such a way that the total
population size remains constant.  Reproduction is also subject to mutation
at some rate $q$ per locus: with probability $q$ the value at any locus in
a sequence changes to a new randomly chosen one when the sequence is
reproduced.

\begin{figure}
\begin{center}
\psfigure{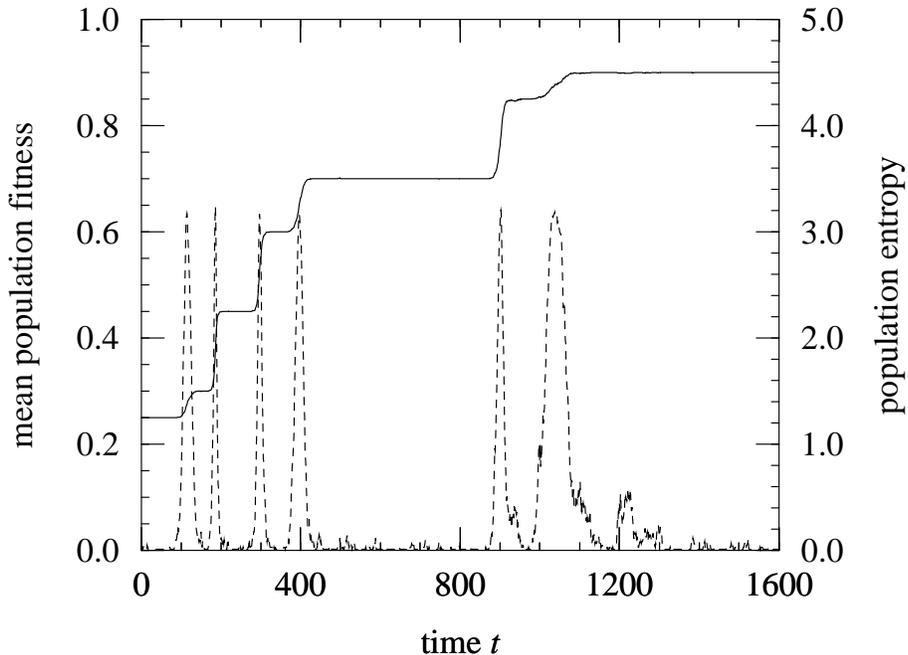}{11.5cm}
\end{center}
\caption{Solid line: the average fitness of a population of $M=1000$
  sequences evolving on a landscape with $N=20$, $K=4$, $A=2$ and $F=2$.
  The mutation rate is $q=5\times10^{-6}$.  Dotted line: the entropy of the
  population, which is a measure of the diversity of sequences present.
\label{fig4}}
\end{figure}

In Figure~\ref{fig4} we show the results of such a simulation for a
population of $M=1000$ sequences with $N=20$ and a mutation rate of
$q=5\times10^{-6}$.  For this value of $q$ the mutation rate is low enough
that the hill-climber approximation used above is reasonable and, as the
figure shows, the epochs visible in the case of the hill-climber appear
also in the average fitness measured in the population simulations (solid
line).  For the simple case of the Royal Road genetic algorithm, van
Nimwegen~\etal~\cite{NCM97b} have studied the epochs extensively and given
a number of analytic and numerical results which may generalize to present
case.\footnote{In fact, the Royal Road model can be reproduced as a special
  case of the present model in which the epistatic interactions between
  loci take a particular structure, being collected in blocks, rather than
  placed randomly between pairs of loci.}

The length of the epochs in Figures~\ref{fig3} and~\ref{fig4} increase, on
average, with increasing fitness.  This behaviour was also seen in the
Royal Road, and occurs because as the fitness increases the number of
structures with higher fitness still dwindles.\footnote{It would do this
  under any circumstances, but the problem is made particularly severe by
  the binomial distribution of fitness values mentioned above, which has an
  exponentially decreasing tail for high fitness.} The length of the epochs
in fact depends on the rate of diffusion of the population across the
neutral network, which in turn depends on $q$, and on the density of
``points of contact'' between the network and other networks of higher
fitness~\cite{Kimura83}.  Another interesting feature of the epochs, also
seen in the Royal Road, is that their average fitness does not correspond
exactly to the fitness of any of the networks.  Typically the average
fitness is a little lower than the fitness of the dominant structure in the
population because deleterious mutants are constantly arising.  Even though
these mutants are selected against, there are at any time enough of them in
the population to make a noticeable difference to the average fitness.
Since the number of possible mutants with lower fitness than the dominant
sequence increases with increasing fitness, it is also possible to get
error threshold effects with increasing fitness~\cite{ES79,SS82}.  As the
fitness increases, there may come a point where the rate at which
deleterious mutants arise in the population exceeds the rate at which they
are suppressed by selection, and at this point further improvement in
fitness becomes impossible.  Thus there may be a dynamical limit on the
fitness of populations, independent of the limit imposed by the structure
of the landscape discussed above.  (This is true of landscapes without
neutral evolution too, though the effect is much more prominent in the
neutral case.)

In Figure~\ref{fig4} we also show the entropy of the population as a
function of time (dotted line).  The entropy is defined as
\begin{equation}
S = -\sum_i p_i \log p_i,
\end{equation}
where $p_i$ is the average probability of finding a particular member of
the population with sequence $i$.\footnote{In fact, calculating this
  entropy for a population of finite size in a large sequence space is not
  a trivial task.  The technical details of the calculation will be
  addressed in more detail in another paper~\cite{NE98}.} In this case, we
see that the entropy is low during the epochs, indicating that the
population is compact, and hence that the strong-selection approximation
made in the case of the random hill-climber applies.  Only during the
selection events themselves, when the population moves to a new neutral
network does the entropy increase.

Simulations similar in spirit to ours have been performed for populations
of tRNAs by Fontana and co-workers~\cite{HSF96,FS87}.  In these simulations
the authors chose a ``target'' structure which was artificially selected
for, and they also observed epochs in the evolution as the population
passed through a succession of increasingly fit structures on its way to
the target.

\section{Discussion}
\label{discuss}
The aim of this work is to study a model of neutral evolution which is
general enough to encompass behaviours typical of other more specific
models which have been employed in the past.  In this way we can reproduce
in a general context the results which have been observed as special cases
and hence investigate the extent to which these results are general
properties of fitness landscapes possessing neutrality, or particular to
the systems in which they were first observed.  In this spirit, we put
forward the following conjectures about the fitness landscapes on which
biological molecules evolve, based on the results of the investigations
outlined in this paper.

\begin{enumerate}
\item The total number of possible structures increases exponentially with
  sequence length.  The exponential constant of this increase appears to be
  approximately numerically equal in both the general model and the only
  specific case in which it has been studied, that of RNA secondary
  structure.
\item There are a large number of structures which correspond to a small
  number of sequences, and a small number of structures which correspond to
  a large number of sequences.  The exact form of the histogram of
  structure frequency, shown in Figure~\ref{fig1}, varies depending on the
  parameters of our model.  However, for certain values of the parameters
  it has a form similar to that seen in RNA studies, whilst for others it
  appears to follow a power law.
\item The ``common'' structures---ones which correspond to a large number
  of sequences---constitute an exponentially decreasing fraction of the
  total number as sequence length increases.  Conversely, however, they
  cover a fraction of the sequence space which tends to unity for long
  sequences.
\item The evolution of populations, at least on short time-scales, is
  dominated by the presence of neutrality.  Neutrality helps the population
  to find structures of high fitness without having to cross fitness
  barriers.  The highest fitness which can be found in this way is closely
  related to the fitness of the highest percolating neutral network, which
  itself depends on the amount of neutrality.  For landscapes with a higher
  degree of neutrality therefore, the population typically reaches a higher
  fitness.
\item The fitness may be limited by error threshold effects, which are
  particularly severe for landscapes of this type, because the size of the
  neutral networks (and hence the ratio of numbers of beneficial and
  harmful mutants) falls exponentially with increasing fitness.
\item Evolution proceeds in jumps separated by ``epochs'' in which the
  fitness appears to change very little.  In fact, the population uses
  these epochs to diffuse across the current neutral network, allowing it
  to search a larger portion of sequence space for beneficial mutations.
\end{enumerate}

\section{Conclusions}
\label{concs}
To conclude, we believe that by studying a simple and general model of a
neutral landscape, we should be able to distinguish properties of specific
systems undergoing neutral selection from properties common to all such
systems.  We have found a number of potential candidates for inclusion in a
list of such common properties.  There are many interesting lines of
investigation which we have not been able to pursue in this short work,
including details of the structure and size of the neutral networks such as
percolation measures and covering radii, details of population dynamics on
these networks including entropy and other statistical measures of the
structure of such populations, calculations of the length of epochs, of the
maximum fitness obtainable on these landscapes, of the effects of error
threshold effects on maximum fitness, and many effects of the variation of
the parameters of the model, particularly the effects of the variation of
the level of epistasis $K$ and the neutrality parameter $F$.  Some of these
questions will be addressed in a forthcoming work.

\section*{Acknowledgements}
The authors would like to thank James Crutchfield, Melanie Mitchell, Erik
van Nimwegen and Paolo Sibani for interesting discussions.  This work was
supported in part by the Santa Fe Institute and DARPA under grant number
ONR N00014--95--1--0975.

\end{document}